\documentclass[12pt]{article}

\usepackage{graphicx}

\usepackage{amsmath,amssymb}
\usepackage{bm}
\usepackage{mathrsfs}
\usepackage{amsfonts}
\usepackage{enumerate}
\usepackage{amsthm}
\usepackage[all,2cell]{xy}
\usepackage{mathrsfs}
\usepackage{mathtools}
\mathtoolsset{showonlyrefs}

\usepackage{tikz-cd}
\usepackage{tikz}
\usetikzlibrary{shapes.multipart}
\usetikzlibrary{patterns,patterns.meta}
\usepackage{xcolor}
\DeclareUnicodeCharacter{0301}{*************************************}

\usepackage[colorlinks=true]{hyperref}
\hypersetup{urlcolor=blue, citecolor=red, linkcolor=blue}

\usetikzlibrary{cd}

\numberwithin{equation}{section}

\usepackage[a4paper, left=25mm, right=25mm, top=30mm, bottom=30mm]{geometry}

\usepackage{marginnote}

\usepackage{imakeidx}
\makeindex[intoc]

\usepackage[nottoc]{tocbibind}
\usepackage[colorinlistoftodos]{todonotes}

\theoremstyle{plain}
\newtheorem{theorem}{Theorem}[section]

\newtheorem{lemma}[theorem]{Lemma}

\theoremstyle{definition}

\newcommand\restr[2]{{
  \left.\kern-\nulldelimiterspace
  #1 
  \right|_{#2}
}}

\newcommand{\R}{\mathbb{R}}
\renewcommand{\d}{\mathrm{d}}

\newcommand{\parder}[2]
{\frac{\partial #1}{\partial #2}}
\newcommand{\D}[2]{\frac{\delta #1}{\delta #2}}

\DeclareMathAlphabet{\mathpzc}{OT1}{pzc}{m}{it}
\def\d{\mathrm{d}}

\def\>{\rangle}
\def\<{\langle}

\let\emph\textbf

\title{{\sffamily Stability analysis of the $(1+1)$-dimensional Nambu-Goto action gas models}}

\author{{\sffamily 
$^{a,c}$ Alfred M. Grundland%
\thanks{e-mail:
   grundlan@crm.umontreal.ca \ ORCID: 0000-0003-4457-7656}\ ,\
   $^{a,b}$ Javier de Lucas%
\thanks{e-mail: javier.de.lucas@fuw.edu.pl \ ORCID: 0000-0001-8643-144X}\ ,\
$^{a,b}$ Bartosz M. Zawora%
\thanks{e-mail:
   b.zawora@uw.edu.pl \ ORCID: 0000-0003-4160-1411}\
}
\\[1ex]
\normalsize\itshape\sffamily
$^a$ Centre de Recherches Mathématiques, Université de Montréal, Succ. Centre-Ville,
\\
\normalsize\itshape\sffamily
CP 6128, Montréal (QC) H3C 3J7, Canada
\\[1ex]
\normalsize\itshape\sffamily
$^b$Department of Mathematical Methods in Physics, University of Warsaw,
\\
\normalsize\itshape\sffamily
ul. Pasteura 5, 02-093 Warszawa, Poland
\\[1ex]
\normalsize\itshape\sffamily
$^c$ Département de Mathématiques et d’Informatique, Université du Québec,
\\
\normalsize\itshape\sffamily
CP 5000, Trois-Rivières (QC) G9A 5H7, Canada
}
\date{{\sffamily }}

\begin{document}

\maketitle

\begin{abstract}
The aim of this paper is to perform a nonlinear stability analysis of the $(1+1)$-dimensional Nambu-Goto action gas models. The energy-Casimir method is employed to discuss in detail the Lyapunov stability of the Chaplygin and Born-Infeld models. Particular solutions are considered and their stability is studied in order to illustrate the application of our results.
\end{abstract}

\bigskip

\bigskip

{\setcounter{tocdepth}{2}
\def\baselinestretch{1}
\small
\def\addvspace#1{\vskip 1pt}
\parskip 0pt plus 0.1mm
\tableofcontents
}

\newpage
\section{Introduction}

The problem of a nonlinear stability of hydrodynamic systems with a Hamiltonian structure was formulated by Arnold et al. in the sixties \cite{Ar65}. Despite its physical relevance, it remained a rather sparse research topic in the following years \cite{BP82, Bl68, Di65}. D. Hold et al. made an important achievement in this field \cite{Ho86, HM85, HM84}. The authors applied the Lyapunov technique adapted by Arnold. They used a class of functionals, the so-called Casimir functionals, which correspond to conserved quantities of the system (i.e. the functionals for which the Poisson bracket commutes). Their construction \cite{HM85} was based on scaling symmetries of the considered systems. It turns out that adding Casimir functionals to the Hamiltonian functional allows for the study of the nonlinear stability of many Hamiltonian systems. The authors present many applications of this technique, establishing new nonlinear stability criteria for equilibrium in various physical systems, including gas dynamics, magnetohydrodynamics, multi-fluid plasma, and Maxwell-Vlasow models in the one-, two-, and three-dimensional cases.

The objective of this paper is to extend the applicability of the energy-Casimir method to systems derived from the Nambu-Goto action \cite{GH04, Ja02}, namely the Chaplygin gas equations and the Born-Infeld equation, for which the stability analysis has never been performed systematically. It is worth understanding that the described approach has a geometric aspect since it is based on a Poisson structure that relates the geometric properties of the symmetries of the system to Casimir functionals. The established Hamiltonian and Poisson structures for these equations constitute a base for the analysis of the geometric features pursued in this study.

The paper by Holm, Marsden, Ratiu, and Weinstein \cite{HM85} studies new nonlinear stability criteria for equilibria in magnetohydrodynamics (MHD), multifluid plasmas, and the Maxwell–Vlasov equations in both two and three dimensions. The authors formulate the energy methods, conserved quantities, and convexity estimates to establish sufficient conditions for stability. Their approach extends the classical energy-Casimir method, providing a framework for analysing the stability of complex fluid and plasma systems.

The Chaplygin gas model and the Born-Infeld model can be derived from the parameterization~-~invariant~Nambu-Goto action, which describes the dynamics of $d$-dimensional branes ($d$-branes) \cite{Ja02}. Specifically, by choosing appropriate parametrizations, the Nambu-Goto action reduces to the equations governing the Chaplygin gas model in the non-relativistic limit and the Born-Infeld model in the relativistic case \cite{Ja02}.

\section{Poisson structures on infinite-dimensional spaces}

This section introduces basic notions and definitions in infinite-dimensional Hamiltonian systems. In particular, it reviews the Hamiltonian formulation of the system of PDEs \cite{BR10, Ko07, Ol93}. It also fixes the notation used hereafter. If not otherwise stated, it is assumed that all structures are smooth and that all spaces are infinite-dimensional Banach spaces.

Hereafter, we assume that $C^\infty(P)$ consists of smooth functions on $P$. In what follows, a {\it local functional} is a functional $F:V\rightarrow C^\infty(\mathbb{R})$ of the form
\[
F(u)=\int f(u^{(r)}(t,x))\d  x,\qquad x\in \mathbb{R},
\]
where \(f\in C^\infty(J^r\mathbb{R}^p)\), the $u^{(r)}$ is the $r$-jet of $u\colon \mathbb{R}^2\rightarrow \R^p$  up to a certain finite order $r$, the integration is over $\R $, and $u\in V$ for a certain Banach space $V$ of functions on $\mathbb{R}^2$ into $\mathbb{R}^p$ so that $F$ is properly defined. From now on,  $(x,t)\in\R^2$, and  $\mathcal{F}$ denotes the space of all local functionals on $V$. The {\it variational derivative} of $F$ at $u$ in the direction $\Delta u\in V$ is the family of elements $\delta F/\delta u^i\in C^\infty(J^r\mathbb{R}^p)$, for $i=1,\ldots,p$, satisfying
\[
\langle \delta F,\Delta u\rangle=\sum^p_{i=1}\int\d x\frac{\delta F}{\delta u^i}(u^{(r)}(t,x))\Delta u^i\,.
\]
Note that, as standard in analysis, the variational derivative does not need to exist in general.

A {\it Poisson bracket} on $\mathcal{A}\subset \mathcal{F}$ is a map $\{\cdot,\cdot\}\colon\mathcal{A}\times\mathcal{A}\rightarrow\mathcal{A}$ that is skew-symmetric and satisfies the Jacobi identity. Recall that when $\mathcal{F}$ is infinite-dimensional, the Poisson bracket is generally defined only in a subclass $\mathcal{A}$ of $\mathcal{F}$ different from $\mathcal{F}$. Moreover, the Leibniz rule is not required since there is no well-defined multiplication between functionals. In any case, we require $\{\cdot,\cdot\}$ to depend bilinearly on the functional derivatives of $F$ so that
$$
\{F,G\}=\sum_{i,j=1}^p\int\frac{\delta F}{\delta u^i}P^{ij}(u)\frac{\delta G}{\delta u^j}dx,
$$
where $P^{ij}(u)$ is an integro-differential operator satisfying appropriate conditions to ensure that $\{\cdot,\cdot\}$ is a Lie bracket and $u=(u^1,\ldots,u^p)$. 

A system of evolution equations 
$$
\frac{\partial u}{\partial t}=X(u),
$$
for a certain mapping $X:V\rightarrow V$  is {\it Hamiltonian} if there is a Poisson bracket $\{\cdot,\cdot\}$ and a functional $H:V\rightarrow C^\infty(\mathbb{R})$ satisfying (see \cite[eq. (EM) and (PB)]{HM85} and \cite{Ol93})
\begin{equation}
\frac{\d}{\d t}F=\{F,H\},\qquad \forall F\in\mathcal{A},
\end{equation}
and conversely \cite{BR10}.  
Poisson brackets on spaces of functions on finite-dimensional manifolds satisfy the Leibniz rule, which ensures that $\{\cdot, H\}$ is related to a vector field. In the infinite-dimensional case, the Poisson bracket is constructed so that the restriction to a local functional $H\in \mathcal{A}$ guarantees the existence of a {\it Hamiltonian vector field}  so that $X_H(F)=\{F, H\}$, for every functional $F\in \mathcal{A}$, see \cite{BR10, Ko07, Ol93}. Here, $X_H$ is understood to be an evolutionary generalized vector field on $V$ \cite[Section 5]{Ol93}. Then, for each functional $F=\int f\d x\in\mathcal{A}$, one has 

\[
X_H(F):=\{F,H\},
\]
and $X_H$ is called a {\it Hamiltonian vector field associated with a Hamiltonian functional $H$}. A functional $C\in\mathcal{A}$ is a {\it Casimir functional}, if 
\[
\{C,F\}=0,\qquad \forall F\in \mathcal{A}.
\]
Within this work, functionals that Poisson commute with Hamiltonian functionals are also called Casimir functionals \cite{HM85}.

\section{Nonlinear stability}

Many authors use the term "stability" in different senses, that is, {\it spectral stability}, {\it linearized stability}, or {\it formal stability} (see \cite{HM85} for details). Here, we will subsequently use the term stability to mean {\it nonlinear stability} in the Lyapunov sense. For conservative systems, it is well-known that even in finite-dimensions, spectral stability is necessary but not sufficient for nonlinear stability. However, neither formal nor linearized stability is required for nonlinear stability \cite{HM85}. 

\subsection{Lyapunov stability}\label{Sec::Stab}

Let us establish some fundamental notions and theorems on the stability of dynamical systems. The Lyapunov stability in the literature is often called nonlinearized stability and this will be the main topic of this section. The methods presented in this section are well-known  \cite{HM85}. However, for the sake of the completeness of our paper, we recall all the steps of the so-called {\it stability algorithm} or {\it energy-Casimir method}. Remarkably, all methods presented here apply in an infinite-dimensional setting.

Consider the system of ODEs, given by
\begin{equation}\label{Eq::NonAutDyn}
     \frac{\d u}{\d t} = X(u)\,,\qquad \forall u\in V\,,
\end{equation}
where $X$ is a vector field on $V$. In particular, we are interested in the case of  $V\subset C^\infty(\mathbb{R}, \mathbb{R}^p)$. In fact, these are functions depending on a space variable $x$. In this sense, solutions to \eqref{Eq::NonAutDyn}  are curves in $V$, which can be considered as mappings $u:\mathbb{R}^2\rightarrow  \mathbb{R}^p$.

A point $u_e\in V$ is an {\it equilibrium point} of \eqref{Eq::NonAutDyn}, or $X$, if $X(u_e)=0$. Furthermore, $u_e$ is {\it stable} if, for every open neighbourhood $\mathcal{U}$ of $u_e$, there exists an open neighbourhood $\mathcal{V}$ such that every solution $u(t,x)$ of \eqref{Eq::NonAutDyn} with initial condition $u(t_0,x)=u_0(x)\in \mathcal{V}$ for some $t_0\in \mathbb{R}$ is such that $u(t,x)$, as a function depending on $x$, is contained in $\mathcal{U}$ for every fixed $t>t_0$. An equilibrium point $u_e\in V$ is {\it unstable} if it is not stable \cite{Vi02}. 

Using the norm on $V$, one says that an equilibrium point $u_e$ is stable if for every $\varepsilon>0$ there exists $\sigma>0$ such that, if $\|u_0(x)-u_e\|<\sigma$, then $\|u(t,x)-u_e\|<\varepsilon$ for every $t\geq t_0$.

\subsection{The energy-Casimir method}

In this subsection, we review the stability algorithm, also known as the {\it energy-Casimir method} \cite{HM85}.

Let $H: V\rightarrow \R$ be a Hamiltonian functional in a family $\mathcal{A}$ such that \[
\frac{dH(u)}{dt}=0
\]
for any solution $u(t,x)$ of \eqref{Eq::NonAutDyn}. Using the Poisson bracket $\{\cdot,\cdot\}$ on $\mathcal{A}$, the equations of motion \eqref{Eq::NonAutDyn} can be rewritten in the following way \cite{HM85}
\[
\frac{d}{dt}F=\{F,H\},\qquad \forall F\in \mathcal{A}\,.
\]
\begin{enumerate}
    \item Therefore, the first step of the algorithm is to find a Hamiltonian functional and compatible Poisson structure.
    \item Then, it is necessary to find a collection of the so-called Casimir functionals $C\in \mathcal{A}$ satisfying
    \[
    \frac{dC(u)}{d t}=0
    \]
    for any solution $u(t,x)$ of \eqref{Eq::NonAutDyn}. Equivalently, Casimir functionals are functionals in $\mathcal{A}$, such that
\[
\{C,H\}=0,
\]
for a Hamiltonian functional $H$. Casimir functionals correspond to the conserved quantities of the considered Hamiltonian system.
\item The next step is to define the {\it extended Hamiltonian functional} $H_C:=H+C\in\mathcal{A}$ and require that, for a given equilibrium point $u_e\in V$,  the first variation of $H_C$ vanishes at $u_e\in V$, namely that $u_e$ be a critical point of $H_C$. At this stage, keeping $C$ as general as possible is important \cite{HM85}.

\item Next, one has to find quadratic forms $Q_1$ and $Q_2$ satisfying
\begin{align}
\label{Eq::Q1form}
    Q_1(\Delta u)&\leq H(u_e+\Delta u)-H(u_e)-\left\langle \delta H(u_e),\Delta u \right\rangle\\
\label{Eq::Q2form}
    Q_2(\Delta u)&\leq C(u_e+\Delta u)-C(u_e)-\left\langle\delta C(u_e),\Delta u\right\rangle\,,
\end{align}
for every $\Delta u:=u-u_e$, where $\delta$ is the functional derivative. Moreover, it is required that
\[
Q_1(v)+Q_2(v)> 0,\qquad \forall v\neq 0.
\]
Therefore, $\|\cdot\|^2=Q_1(\,\cdot\,)+Q_2(\,\cdot\,)$ defines a norm. Assuming that $H_C\in\mathcal{A}$ is continuous in this norm at $u_e\in V$, it follows that $u_e\in V$ is a stable equilibrium point.
\end{enumerate}

The above algorithm can be expressed via the following theorem. The proof can be found in \cite[Section 2]{HM85}.

\begin{theorem}
\label{Th::energy-Casimir}
    Let $(\mathcal{A},\{\cdot,\cdot\})$ be a Poisson Banach space and let $H\in\mathcal{A}$ be a Hamiltonian functional such that $\dot{H}(u)=0$ for any solution $u\in V$ of \eqref{Eq::NonAutDyn}. Additionally, let $C\in\mathcal{A}$ denote a Casimir functional, i.e. $\{C,\,\cdot\,\}=0$ and let $H_C:=H+C\in\mathcal{A}$ be continuous in the norm defined by $\|\cdot\|^2=Q_1(\,\cdot\,)+Q_2(\,\cdot\,)$ which admits a critical point at $u_e\in V$. Moreover, assume that there exist quadratic forms $Q_1$ and $Q_2$ satisfying
    \begin{align}
    \label{Eq::Q1Q2}
        Q_1(\Delta u)&\leq H(u_e+\Delta u)-H(u_e)-\langle\delta H(u_e),\Delta u\rangle,\\
        Q_2(\Delta u)&\leq C(u_e+\Delta u)-C(u_e)-\langle\delta C(u_e),\Delta u\rangle,
    \end{align}
    and
    \begin{equation}
    \label{Eq::Norm}
    \|v\|^2:=Q_1(v)+Q_2(v)> 0,\qquad \forall v\neq 0.
    \end{equation}
    Then, indeed $\|\cdot\|$ is a norm and $u_e\in V$ is a stable equilibrium point of \eqref{Eq::NonAutDyn}.
\end{theorem}

The following lemma gives sufficient conditions to ensure that a functional $H_C\in \mathcal{A}$ is continuous in $\|\cdot\|$.

\begin{lemma}
\label{Lemm::energy-Casmir_cont}
    Let $H\in\mathcal{A}$ be a Hamiltonian functional and suppose that there exist constants $a_1,a_2\in \R$ such that
    \begin{align}
    \label{Eq::ContCond1}
        H(u_e+\Delta u)-H(u_e)-\langle\delta H(u_e), \Delta u\rangle &\leq a_1 \|\Delta u\|^2\\
        \label{Eq::ContCond2}
        C(u_e+\Delta u)-C(u_e)-\d \langle C(u_e), \Delta u\rangle &\leq a_2 \|\Delta u\|^2,
    \end{align}
    where $\|\cdot\|:=Q_1(\,\cdot\,)+Q_2(\,\cdot\,)$ is defined as in Theorem \ref{Th::energy-Casimir}, for every $\Delta u$. Then, $H_C=H+C\in\mathcal{A}$ is continuous in $\|\cdot\|$.
\end{lemma}

\section{Chaplygin gas model}
This section discusses the energy-Casimir algorithm for the Chaplygin gas model in \((1+1)\)-dimensions. First, the Lagrangian and Hamiltonian formulations are presented. Then, the energy-Casimir stability algorithm is carried out step-by-step. In the last subsection, particular solutions of the Chaplygin gas model are analyzed. 
\subsection{Lagrangian formulation of the Chaplygin gas model}
The Lagrangian setting of the $(1+1)$-dimensional Chaplygin gas model is as follows
\[
L=\int\left(\theta\frac{\partial\rho}{\partial t}-\rho\frac{(\nabla\theta)^2}{2}-\frac{\lambda}{\rho}\right)\d x\,,\qquad \lambda\in \R\,,
\]
where $\theta$ and $\rho$ are the velocity potential and the density of the system, respectively \cite[p. 32]{Ja02}. Hence, $\rho$ is assumed to be greater than zero everywhere. Recall that the integration is performed over $\R$. The corresponding Euler–Lagrange equations, read
\begin{equation}
\label{Eq::ChaplyginHamEqs2}
\begin{aligned}
        \frac{\partial\rho}{\partial t}&=-\frac{\partial}{\partial x}\left(\frac{\partial\theta}{\partial x}\rho\right),\\
        \frac{\partial p}{\partial t}&=-\frac{\partial}{\partial x}\left(\frac{(\frac{\partial\theta}{\partial x})^2}{2}-\frac{\lambda}{\rho^2}\right)\,.
\end{aligned}
\end{equation}
Then, from \eqref{Eq::ChaplyginHamEqs}, by eliminating $\rho$, one gets the so-called {\it Chaplygin equation} given by
\begin{equation}
\label{Eq::ChaplyginEQ}
\frac{\partial}{\partial t}\left(\frac{1}{\sqrt{\frac{\partial\theta}{\partial t}+\frac{1}{2} p^2}}\right)+\frac{\partial}{\partial x}\left(\frac{\frac{\partial\theta}{\partial x}}{\sqrt{\frac{\partial\theta}{\partial t}+\frac{1}{2}p^2}}\right)=0\,,
\end{equation}
The relation between $\theta$ and $p$ is that $\frac{\partial\theta}{\partial x}=p$, where $p$ is the momentum \cite[Section 6]{Ja02}.
\subsection{Hamiltonian formulation of the Chaplygin gas model}
The Hamiltonian functional $H\in\mathcal{A}$ for the $(1+1)$-dimensional Chaplygin gas model is given by
\[
H=\int \left(\frac{1}{2}\rho p^2+\frac{\lambda}{\rho}\right)\d x\,,\qquad \lambda\in \R\,.
\]
where $\rho$ and $p$ denote density and momentum, respectively \cite{Ja02}. The associated equations of motion are of the form
\begin{equation}
\label{Eq::ChaplyginHamEqs}
\begin{aligned}
        \frac{\partial\rho}{\partial t}&=-\frac{\partial}{\partial x}(p\rho),\\
        \frac{\partial p}{\partial t}&=-\frac{\partial}{\partial x}\left(\frac{p^2}{2}-\frac{\lambda}{\rho^2}\right)\,.
\end{aligned}
\end{equation}

\subsection{Stability of the \texorpdfstring{\((1+1)\)}--dimensional Chaplygin gas model}

In this section, the stability algorithm for a Chaplygin gas model is presented. 

Recall that the Hamiltonian functional $H\in\mathcal{A}$ for a Chaplygin gas model reads
\[
H=\int \left(\frac{1}{2}\rho p^2+\frac{\lambda}{\rho}\right)\d x,
\]
The Poisson structure related to the Chaplygin gas model has the form \cite{BR10}
\begin{equation}
\label{Eq:PoissonChapGas}
\{F,G\}=-\int \left(\frac{\delta F}{\delta \rho}\nabla \frac{\delta G}{\delta p}+\frac{\delta F}{\delta p}\nabla\frac{\delta G}{\delta\rho}\right)\d x \,,
\end{equation}
for any functionals $F=\int f(p,\rho)\d x$ and $G=\int g(p,\rho)\d x$.

Note that this Poisson bracket reproduces the Hamilton equations associated with the $(1+1)$-dimensional Chaplygin gas model \eqref{Eq::ChaplyginHamEqs}, namely
\begin{equation}
\label{Eq::EqMotionChaplyginGas}
    \begin{aligned}
        \frac{\partial p}{\partial t}&=-\frac{\partial}{\partial x}\left(\frac{p^2}{2}-\frac{\lambda}{\rho^2}\right),\\
        \frac{\partial \rho}{\partial t}&=-\frac{\partial}{\partial x}\left(p\rho\right),
    \end{aligned}
\end{equation}
for some constant $\lambda>0$. Equations \eqref{Eq::EqMotionChaplyginGas} give rise to a Hamiltonian vector field of the form
\begin{equation}
\label{Eq::HamVecChaplyginGas}
X_H=-\left(\frac{\partial}{\partial x}\left(\frac{p^2}{2}-\frac{\lambda}{\rho^2}\right)\right)\parder{}{p}-\left(\frac{\partial}{\partial x}\left(p\rho\right)\right)\parder{}{\rho}.
\end{equation}
One can check that the Hamiltonian vector field \eqref{Eq::HamVecChaplyginGas} has an equilibrium point for some $u=(p,\rho)$ satisfying $p\rho=\sqrt{2\lambda}$, namely $X_H(u)=0$.

Recall, one has to find a functional $C\in\mathcal{A}$ such that $\{C,H\}=0$. The functional of the form
\[
C=\int f\left(\frac{\rho}{p}\right)\d x\,,
\]
where $f$ is arbitrary, with the additional condition that $p\rho=\sqrt{2\lambda}$, satisfies
\begin{multline*}
    \{C,H\}=\int\left( \frac{\delta C}{\delta \rho}\nabla\left(p\rho\right)+\frac{\delta C}{\delta p}\nabla\left(\frac{p^2}{2}-\frac{\lambda}{\rho^2}\right)\right)\d x\\=\int\left(\frac{\delta C}{\delta p}\left(p\nabla p+\frac{2\lambda}{\rho^3}\nabla \rho\right)+\frac{\delta C}{\delta\rho}\left(p\nabla \rho+\rho\nabla p\right)\right)\d x\\=\int\left(\nabla p\left(p\frac{\delta C}{\delta p}+\rho\frac{\delta C}{\delta\rho}\right)+\nabla\rho\left(\frac{2\lambda}{\rho^3}\D{C}{p}+p\D{C}{\rho}\right)\right)\d x\\=\int\nabla\rho\left(\frac{2\lambda}{\rho^3}\D{C}{p}+p\D{C}{\rho}\right) \d x=\int\dot{f}\left(1-\frac{2\lambda}{\rho^2p^2}\right) \d x=0.
\end{multline*}
Proceeding to the third step in the energy-Casimir method, let us assume that $p\rho=\sqrt{2\lambda}$. Define the functional
\[
H_C=\int\left(\frac{1}{2}\rho p^2+\frac{\lambda}{\rho}+f\left(\frac{p}{\rho}\right)\right) \d x,
\]
and impose the condition
\[
\delta H_C(u_e)=0\,,
\]
where $u_e$ is an equilibrium point. Then, from $p\rho=\sqrt{2\lambda}$, it follows that
\begin{multline*}
    \delta H_C(u_e)=\int \left(\left(p_e\rho_e -\frac{\rho_e}{p_e^2}\dot{f}_e\right)\delta p + \left(\frac{1}{2}p^2_e-\frac{\lambda}{\rho^2_e}+\frac{1}{p_e}\dot{f}_e\right)\delta\rho\right)\d x\\=\sqrt{2\lambda}\int\left(1-\frac{2\dot{f}}{p_e^3}\right)\delta p\,\d x=0.
\end{multline*}
The above expression equals zero if, for instance, $\dot{f}=1$ and $p_e=\sqrt[3]{2}$. Therefore, the Casimir functional $C$ has the form
\[
C=\int \frac{\rho}{p}\d x\,,
\]
and the equilibrium point is $(p_e,\rho_e)=(\sqrt[3]{2},\frac{\sqrt{2\lambda}}{\sqrt[3]{2}})$.
The next step is to find the quadratic forms $Q_1$ and $Q_2$ satisfying \eqref{Eq::Q1form} and \eqref{Eq::Q2form}. Hence, the right-hand side of \eqref{Eq::Q1form} gives
\begin{multline}
\label{Eq::S1Q1}
   \int\left( \frac{1}{2}(\Delta\rho+\rho_e)(\Delta p+p_e)^2+\frac{\lambda}{\Delta\rho+\rho_e}-\frac{1}{2}p_e^2\rho_e-\frac{\lambda}{\rho_e}-\left(\rho_ep_e\Delta p+\left(\frac{1}{2}p_e^2-\frac{\lambda}{\rho_e^2}\right)\right)\right)\d x\\=\int\left(\frac{1}{2}\left((\Delta\rho+\rho_e)(\Delta p+p_e)^2-\rho_ep_e^2-2\rho_ep_e\Delta p-p_e^2\Delta\rho\right)+\lambda\left(\frac{1}{\Delta\rho+\rho_e}-\frac{1}{\rho_e}+\frac{\Delta\rho}{\rho_e^2}\right)\right)\d x\\=\int\left(\frac{1}{2}\left( \Delta p^2(\Delta\rho+\rho_e)+2p_e\Delta p\Delta \rho\right)+\lambda\frac{\Delta\rho^2}{\rho_e^2(\Delta\rho+\rho_e)}\right)\d x.
\end{multline}
Keeping in mind that $\Delta u=u-u_e$, one gets that equation \eqref{Eq::S1Q1} reduces to
\begin{multline}
\label{Eq::S2Q1}
\int\left(\frac{1}{2}\left(\Delta p^2\rho+2p_e\Delta p\Delta \rho\right)+\lambda\frac{\Delta\rho^2}{\rho_e^2\rho}\right)\d^nx\\=\int\left(\frac{1}{2}\left(\frac{\sqrt{2\lambda}}{p}\Delta p^2-2\sqrt{2\lambda}\frac{\Delta p^2}{p}\right)+\sqrt{2\lambda}\frac{\Delta p^2}{2p}\right)\d^nx=0.
\end{multline}
Therefore, $Q_1=0$. Note that if $Q_2>0$, then one can construct the norm required in the energy-Casimir method given by equation \eqref{Eq::Norm}. Hence, similarly for \eqref{Eq::Q2form}, the right hand side reads
\[
\int\left(\frac{\Delta p+p_e}{\Delta\rho+\rho_e}-\frac{p_e}{\rho_e}-\frac{\Delta p }{\rho_e}+\frac{p_e}{\rho_e^2}\Delta\rho\right)\d x=\int\left(\frac{1}{\rho_e^2(\Delta\rho+\rho_e)}\left(p_e\Delta\rho^2-\rho_e\Delta p\Delta \rho\right)\right)\d x\,.
\]
Taking into account that $p\rho=\sqrt{2\lambda}$, we get
\[
\int\left(\frac{1}{\rho_e^2(\Delta\rho+\rho_e)}\left(p_e\Delta\rho^2-\rho_e\Delta p\Delta \rho\right)\right)\d x=\int\left(\frac{\sqrt{2\lambda}}{\rho_e\rho}\left(\frac{1}{\rho_e}+\frac{1}{\rho}\right)\Delta\rho^2\right)\d x\,.
\]
Recall that $\rho(x,t)>0$. Additionally, let us assume that there exist constants such that $\alpha:=\inf_{(x,t)\in\R^{2}}\rho(x,t)>0$ and $\beta:=\sup_{(x,t)\in\R^{2}}\rho(x,t)>0$.
Then, taking 
\[
Q_2(\Delta\rho)=\int\left(\frac{\sqrt{2\lambda}}{\rho_e\beta}\left(\frac{1}{\rho_e}+\frac{1}{\beta}\right)\Delta\rho^2\right)\d x\,,
\]
one gets that $Q_2$ is positive definite, satisfies \eqref{Eq::Q2form}, and defines a norm on $P$, given by
\[
\|v\|^2=Q_2(v)>0,\qquad\forall v\neq 0\,.
\]
Next, one has to check whether $H_C$ is continuous in the norm $\|\cdot\|$ at $u_e$. Therefore, it is sufficient to check if the conditions \eqref{Eq::ContCond1} and \eqref{Eq::ContCond2} hold. Therefore,
\[
H(\Delta u +u_e)-H(u_e)-\left\langle\delta H(u_e),\Delta u\right\rangle=0\leq\|\Delta u\|^2\,,
\]
and
\[
C(\Delta u +u_e)-C(u_e)-\left\langle\delta C(u_e),\Delta u\right\rangle=\int\left(\frac{\sqrt{2\lambda}}{\rho_e\rho}\left(\frac{1}{\rho_e}+\frac{1}{\rho}\right)\Delta\rho^2\right)\d x\leq a_2\|\Delta u \|^2\,,
\]
for $a_2:=\frac{\beta^2}{\alpha^2}$. Therefore, conditions \eqref{Eq::ContCond1} and \eqref{Eq::ContCond2} hold and Lemma \ref{Lemm::energy-Casmir_cont} yields that $H_C$ is continuous in $\|\cdot\|$. Then, by Theorem \ref{Th::energy-Casimir} one gets that $u_e=(p_e,\rho_e)=(\sqrt[3]{2},\frac{\sqrt{2\lambda}}{\sqrt[3]{2}})$ is a stable equilibrium point on $p\rho=\sqrt{2\lambda}$ in the Lyapunov sense.

\section{Born-Infeld model}

This section presents the Lagrangian and Hamiltonian formulations of the Born-Infeld model and applies the energy-Casimir method in its analysis.

\subsection{Lagrangian formulation of the Born-Infeld model}

The Lagrangian of the $(1+1)$-dimensional Born-Infeld model \cite[Section 6]{Ja02}, reads 
\[
L=\int \left(\theta \frac{\partial\rho}{\partial t}-\sqrt{(\rho^2c^2+a^2)(c^2+(\nabla\theta)^2)}\right)\d x,
\]
where $a$ is an interaction strength and $c$ the speed of light. The corresponding equations of motion are given by
\begin{equation}
    \label{}
    \begin{aligned}
        \frac{\partial \rho}{\partial t}+\frac{\partial}{\partial x}\left(\frac{\partial\theta}{\partial x}\sqrt{\frac{\rho^2c^2+a^2}{c^2+(\frac{\partial\theta}{\partial x})^2}}\right)=0\,,\\
        \frac{\partial\theta}{\partial t}+\rho c^2\sqrt{\frac{c^2+(\frac{\partial\theta}{\partial x})^2}{\rho^2c^2+a^2}}=0\,.
    \end{aligned}
\end{equation}
In particular, $\partial \theta/\partial x=a/\rho(x)$ and $\theta=\int a/\rho(x)dx-t$ satisfy these conditions. Additionally, to ensure the necessary bounds on $\rho$, one can consider $\rho=2+\cos(x)$, for instance. 
Then, after the evaluation of $\rho$ it terms of $\theta$, one gets
\begin{equation}\label{eq:BornInfeld}
\frac{1}{c^2}\frac{\partial}{\partial t}\left(\frac{\frac{\partial\theta}{\partial t}}{\sqrt{c^2-\frac{1}{c^2}(\frac{\partial \theta}{\partial t})^2+p^2}}\right)-\nabla\left(\frac{p}{\sqrt{c^2-\frac{1}{c^2}(\frac{\partial\theta}{\partial t})^2+p^2}}\right)=0,
\end{equation}
where $p=\nabla \theta=\frac{\partial \theta}{\partial x}$. Equation \ref{eq:BornInfeld} is known as the {\it Born-Infeld} equation.

\subsection{Hamiltonian formulation of the Born-Infeld model}

The Hamiltonian of the Born-Infeld model \cite[Section 6]{Ja02} is given by
\begin{equation}
\label{Eq::BIHam}
H=\int  \left(\sqrt{\rho^2c^2+a^2}\sqrt{c^2+p^2}\right)\d x\,,
\end{equation}
where $a,c\in\R$. 
The corresponding Hamiltonian equations \cite[Section 6]{Ja02} read
\begin{align}
\label{Eq::BIEq}
        \frac{\partial\rho}{\partial t}=&-\nabla\left(p\frac{\sqrt{\rho^2c^2+a^2}}{\sqrt{c^2+p^2}}\right),\\
        \frac{\partial p}{\partial t}=&-\nabla\left(\rho c^2\frac{\sqrt{c^2+p^2}}{\sqrt{\rho^2c^2+a^2}}\right)\,.
\end{align}
Note that taking the limit $c\rightarrow \infty$ in \eqref{Eq::BIEq} one obtains the Chaplygin gas model, with $\lambda=\frac{a^2}{2}$.

\subsection{Stability algorithm for the Born-Infeld model}

In this subsection, we apply the stability algorithm to the Born-Infeld model. Additionally, a particular solution is analyzed.

The Poisson structure that reproduces the Hamiltonian equations associated with the $(1+1)$-dimensional Born-Infeld model is the same as for the Chaplygin gas model, namely
\[
\{F,G\}=-\int \left(\frac{\delta F}{\delta \rho}\nabla \frac{\delta G}{\delta p}+\frac{\delta F}{\delta p}\nabla\frac{\delta G}{\delta\rho}\right) \d x\,.
\]
Note that this Poisson bracket recovers the equation of motion \eqref{Eq::BIEq}. The Hamiltonian vector field has the following form
\[
X_H=-\nabla\left(p\frac{\sqrt{\rho^2c^2+a^2}}{\sqrt{c^2+p^2}}\right)\frac{\partial}{\partial \rho}-\nabla\left(\rho c^2\frac{\sqrt{c^2+p^2}}{\sqrt{\rho^2c^2+a^2}}\right)\frac{\partial}{\partial p}\,.
\]
From now on, assume that $c=a=1$. Then, the equilibrium point of $X_H$ is a point $u=(p,\rho)$, where $p\rho=1$, i.e. $X_H(u_e)=0$. Then, one needs a functional $C$ that satisfies $\{C,H\}=0$. First, let us restrict ourselves to the case when $p\rho=1$. Then it follows that $\{F, H\}=0$ for every functional $F$.

The second step of the algorithm is as follows. Define $H_C:=H+C$, for some functional $C$. Let us assume that
\[
C(\rho)=-\frac{1}{2}\int \left(\frac{1}{\rho}+\rho\right)\d x.
\]
Then, it is required that $\delta H_C(u_e)=0$. Hence, taking into account the constraint $p\rho=1$, one gets
\begin{align}
    \delta H_C&=\int \left(\left(p\sqrt{\frac{\rho^2+1}{p^2+1}}+\frac{\delta C}{\delta p} \right)\delta p+\left( \rho\sqrt{\frac{p^2+1}{\rho^2+1}}+\frac{\delta C}{\delta \rho}\right)\delta\rho\right)\d x\\&=\int \left(-\frac{1}{\rho^2}\delta \rho+\left(1+\frac{\delta C}{\delta \rho}\right)\delta\rho\right)\d x=\int \left(-\frac{1}{\rho^2}+\frac{\delta C}{\delta \rho}+1\right)\delta\rho\,\d x\\&=\frac{1}{2}\int \left(1-\frac{1}{\rho^2}\right)\delta\rho\,\d x\,.
\end{align}
The above expression vanishes if $\rho=1$, hence $p=1$. Note that $C$ takes the form of a Hamiltonian with the constraint given by $p\rho =1$.

Then, the next step of the energy-Casimir algorithm requires us to find quadratic forms $Q_1$ and $Q_2$ satisfying \eqref{Eq::Q1form} and \eqref{Eq::Q2form}. However, since $C$ is proportional to $H$, we need to find only $Q_1$. Then, restricting to $p\rho=1$, we get
\begin{multline*}
H(u_e+\Delta u)-H(u_e)-\left\langle\delta H(u_e),\Delta u\right\rangle=\int\left(\rho_e+\Delta\rho +\frac{1}{\rho_e+\Delta\rho}-2\right)\d x\\=\int\left(\frac{\rho_e^2+\Delta\rho^2+2\rho_e\Delta\rho-1-2\Delta\rho}{\rho}\right)\d x=\int\frac{\Delta\rho^2}{\rho}\d x\,,
\end{multline*}
where $\Delta\rho=\rho-\rho_e$. Assuming that there exist constants $\beta=\sup_{(t,x)\in\R^2}\rho(t,x)>0$ and $\gamma=\inf_{(t,x)\in\R^2}\rho(t,x)>0$, it follows that
\[
Q_1(\Delta\rho)=\int\frac{\Delta\rho^2}{\beta}\d x.
\]
Then, condition \eqref{Eq::Q1form} holds. To determine stability, one has to ensure that $H_C$ is continuous in $\|\cdot\|$ defined through $\|\cdot\|^2=Q_1(\,\cdot\,)$. In other words, the assumptions of Lemma \ref{Lemm::energy-Casmir_cont} must be satisfied. Note that
\[
H(u_e+\Delta u)-H(u_e)-\left\langle\delta H(u_e),\Delta u\right\rangle\leq a\, Q_1(\Delta u),
\]
holds for $a:=\beta/\gamma$. Thus, according to Theorem \ref{Th::energy-Casimir}, a point $(p,\rho)=(1,1)$ is a stable equilibrium point on $p\rho=1$ in the Lyapunov sense.

\section{Stability analysis of particular solutions}

In this section, the nonlinear stability of particular solutions of the Chaplygin gas model and Born-Infeld model are studied to illustrate the applications of our theory.

\subsection{Chaplygin gas model}

Consider particular solution of the Chaplygin gas model in $(1+1)$-dimensions \eqref{Eq::ChaplyginHamEqs} of the form
\begin{equation}
    \begin{aligned}
        \theta&=\sin x+2x-t\,,\\
        \rho&=\frac{\sqrt{2\lambda}}{\cos x+2}\,.        
    \end{aligned}
    \qquad\qquad(t,x)\in \R^2\,.
\end{equation}
Then, the momentum $p=\frac{\partial\theta }{\partial x}$ reads
\[
p=\cos x+2\,.
\]
Note that the condition $p\rho=\sqrt{2\lambda}$ is satisfied and $\frac{\sqrt{2\lambda}}{3}\leq \rho(t,x)\leq \sqrt{2\lambda}$ for every $(t,x)\in\R^2$. Therefore, a point $(p_e,\rho_e)=(\sqrt[3]{2},\frac{\sqrt{2\lambda}}{\sqrt[3]{2}})$ is a stable equilibrium point. 

\subsection{Born-Infeld model}

Consider particular solution of the Born-Infeld model in $(1+1)$-dimensions \eqref{Eq::BIHam} of the form
\begin{equation}
    \begin{aligned}
        \theta&=\sin x+2x-t\,,\\
        \rho&=\frac{1}{\cos x+2}\,.        
    \end{aligned}
    \qquad\qquad(t,x)\in \R^2\,.
\end{equation}
The corresponding momentum $p = \frac{\partial \theta}{\partial x}$ is given by  
\[
p = \cos x + 2.
\]  
Moreover, the condition $ p \rho = 1$ holds, and the function $\rho$ satisfies $\frac{1}{3} \leq \rho(t,x) \leq 1$  for all $(t,x) \in \mathbb{R}^2$. Consequently, the point $(p_e, \rho_e)= (1,1)$ is a stable equilibrium point.

\addcontentsline{toc}{section}{Acknowledgements}
\section*{Acknowledgements}
This work was partially supported by A.M. Grundland's grant from NSERC of Canada. J.
de Lucas acknowledges a Simons–CRM professorship funded by the Simons Foundation and
the Centre de Recherches Mathématiques (CRM) of the Université de Montréal. B.M. Zawora acknowledges funding from the IDUB
Mikrogrant program to accomplish a research stay at the CRM and thanks the Centre de Recherches Mathématiques (CRM) of the Université de Montréal and the CRM Simons Foundation for their hospitality.

\end{document}